\documentclass[12pt,a4paper]{iopart}
\usepackage{iopams}
\usepackage{graphicx}
\usepackage[usenames]{color}
\usepackage[normalem]{ulem}

\newcommand{\be}{\begin{equation}}
\newcommand{\ee}{\end{equation}}
\newcommand{\ba}{\begin{eqnarray}}
\newcommand{\ea}{\end{eqnarray}}
\newcommand{\baa}{\begin{eqnarray}}
\newcommand{\eaa}{\end{eqnarray}}
\newcommand{\ed}{\end{document}}

\renewcommand{\baselinestretch}{1.2}
\date{\today}

\begin{document}%large

\title[Dynamics of trapped interacting vortices in BECs: Role of breathing degree]%
{Dynamics  of trapped interacting vortices in Bose-Einstein condensates: Role of breathing degree of freedom}
\author{Katsuhiro Nakamura$^{1,2}$, Doniyor Babajanov$^{3}$, Davron Matrasulov$^{3}$, Michikazu Kobayashi$^{4}$, and Paulsamy Muruganandam$^{5}$}
\address{$^{1}$Faculty of Physics, National University of Uzbekistan, Vuzgorodok, Tashkent 100174,Uzbekistan}
\address{$^{2}$Department of Applied Physics, Osaka City University, Osaka 558-8585, Japan}
\address{$^{3}$Turin Polytechnic University in Tashkent, 17 Niyazov Str., Tashkent 100095, Uzbekistan}
\address{$^{4}$Department of Physics, Kyoto University, Kyoto 606-8502, Japan}
\address{$^{5}$Department of Physics, Bharathidasan University, Tiruchirappali 620024, India}

\begin{abstract}
With use of a variational principle, we investigate a role of breathing width degree of freedom  in the effective theory of interacting vortices in a trapped single-component Bose-Einstein condensates in 2
dimensions under the strong repulsive cubic nonlinearity.
As for the trial function, we choose a product of two vortex functions assuming  a pair interaction and employ the amplitude form of each vortex function in the  Pad\'e approximation which accommodates a hallmark of the vortex core. We have obtained 
Lagrange equation for the interacting vortex-core coordinates coupled with the time-derivative of width 
and also its Hamilton formalism by having recourse to a non-standard Poisson bracket. 
By solving the Hamilton equation, we find rapid radial breathing oscillations superposed on the slower rotational motion of vortex cores, consistent with numerical solutions of Gross-Pitaevskii equation. In higher-energy states of 
2 vortex systems, the breathing width degree of freedom plays role of a kicking in the kicked rotator and generates chaos with a structure of sea-urchin needles. 
Byproduct of the present variational approach includes: (1) the charge-dependent logarithmic inter-vortex interaction multiplied with a pre-factor which depends on the scalar product of a pair of core-position vectors; (2) the charge-independent short-range repulsive inter-vortex interaction and spring force.
\end{abstract}
\maketitle

\section{Introduction}\label{sec-introduction}

Trapped single-component \cite{rf:nee,rf:frei,torr,nava} and  multi-component \cite{rf:tha,rf:pap,rf:tojo,Dant} Bose-Einstein condensates
(BECs) have been
realized experimentally, giving a good candidate in which to study theoretically
the dynamics of vortices.  In particular, recent attention is paid to small vortex clusters
\cite{rf:nee,rf:frei,torr,nava,Sem}.
While dynamics of the macroscopic wave function of  BECs is described by the nonlinear Schr{\"o}dinger equation or Gross-Pitaevskii equation (GPE) as in equation~(\ref{eq-GP}) below,
the trapping potential plays a nontrivial role. 

For trapped single-component  BECs, the presence of the breathing mode of width, i.e, collective radial oscillation of the spherically- or circularly-confined region is exhibited both theoretically and experimentally\cite{Stri,Jin,Mew,fett1,fett2}. On the other hand, the effective dynamics of interacting vortices in trapped BECs with strong nonlinearity is recently proposed and numerically analyzed
\cite{torr,nava,Kou,Kol}, but
little theoretical attention has been paid to how to couple the core-coordinates of interacting vortices  with the breathing width degree of freedom.

In our previous paper \cite{nakam},
we developed an effective theory of vortex dynamics in two-dimensional(2-d) multi-component BEC in the
harmonic trap and with strong nonlinearity in the case that each component has a "single vortex".
There we chose the amplitude form of each vortex function in the Pad\'e approximation which accommodates a hallmark of the vortex core. The important finding there was the nonzero inertia (or momentum) of vortices suggested in \cite{qmc2008}, which was distinct from the inertia-less vortices widely used in the usual hydrodynamics
\cite{rf:Neu,rf:Afta,rf:aref}. We also recognized little active role of the breathing degree of freedom in single vortex dynamics. However, these conclusions would become questionable in the case that each component of BECs will accommodate "plural number of vortices", because of the strong phase correlation among vortices.

In this paper, we apply the standard variational principle to the interacting vortices in 2-d non-rotating trapped
single-component BECs with strong cubic nonlinearity and elucidate an outstanding  role of the breathing degree of freedom which is coupled with
vortex degree of freedom (i.e., core coordinates).

This paper is organized as follows. In section \ref{sec-model},
we apply the variational method with use of
a regularized function for a pair of vortices which includes the width degree of freedom as a collective coordinate, derive Lagrange equations for vortex cores and width, and construct their Hamiltonian formalism.
In section \ref{sec-some-vortices}, solving numerically the above Hamilton equation, we show novel effects of breathing oscillations on vortices, which will be verified by numerical iteration of GPE. We  proceed to dynamics of two vortices in both cases of identical and different vortex charges. There we see emergence of
chaos in higher energies in trapped two-vortex systems under the breathing dynamics. 
Section \ref{sec-conclusion} is
devoted to conclusions and discussions.

\section{Nonlinear dynamics of interacting vortices coupled with breathing degree of freedom}\label{sec-model}

We consider a 2-d trapped non-rotating single-component BEC with
two vortices described by GPE. The macroscopic wave
function $\Phi(t,{\bf r})$ with ${\bf r}=(x,y)$
satisfies the equation: 
\begin{equation}
 i\frac{\partial}{\partial t}\Phi(t,{\bf r})
= \left[ -\nabla^2+V({\bf r})+g|\Phi(t,{\bf r})|^2\right]\Phi(t,{\bf r}).
\label{eq-GP}
\end{equation}
Here the normalization condition is defined by
$\int |\Phi(t,{\bf r})|^2 d^2{\bf r}=1$ after a proper rescaling of $\Phi$.
The effect of trapping is expressed by
$V({\bf r})={\bf r}^2$.

To derive equation~(\ref{eq-GP}), we consider the ultracold pancake-shaped condensate 
in the $xy$-plane with the anisotropic angular frequencies $\omega_z$
and $\omega_r$ with $\omega_r \ll  \omega_z$ for the harmonic trap,
and neglect the degrees of freedom along $z$ direction of the wave
function, i.e., $\Phi(t,x,y,z) \sim \frac{\Phi(t,x,y)}{\sqrt{a_z}}$
with the typical size  $a_z = \sqrt{\frac{\hbar}{m \omega_z}}$ of the condensate along $z$-axis. By using
the confining length $a_r = \sqrt{\frac{\hbar}{m \omega_r}}$ and oscillation
period $\tau = \omega_r^{-1}$, space coordinates are scaled by $a_r$,
time by $2 \tau$, wave function by $\frac{1}{a_r}$, and nonlinearity by
 $\frac{\hbar^2}{2m}$. The nonlinearity coefficient is defined by
$g = \frac{8 \pi \mathcal{N} a}{a_z}$ with $a$ the positive $s$-wave scattering length
for binary collisions and $\mathcal{N}$ the particle number. We shall investigate the case of $g \gg 1$,
namely, strong repulsive nonlinearity.
In typical experimental setups, i.e., $\omega_z \sim 1000$Hz, $\omega_r \sim 100$Hz,
and $\mathcal{N} \sim 10^5$, a typical Thomas-Fermi radius in a $z$-direction is
$R_z = \sqrt[5]{\frac{15 \mathcal{N} a a_z}{a_r^4}} \sim 2.5 a_z$ for $^{87}$Rb atoms: $m \sim 1.5 \times 10^{-25}$kg and
$a \sim 110 a_{\mathrm{B}}$ with the Bohr radius $a_{\mathrm{B}}$.
Therefore, the thickness of condensate cloud in a $z$-direction is the same order as $a_z$ and the assumption for the
wave function $\Phi(t,x,y,z) \sim \frac{\Phi(t,x,y)}{\sqrt{a_z}}$ is justified within these experimental
setups. The nonlinearity is also estimated as $g \sim 10^4$. 

As for the trial vortex function (TVF), we choose a product of two vortex functions assuming a pair interaction and employ the amplitude form of each vortex function on the basis of Pad\'e approximation~\cite{rf:Igor,natali}. 
TVF is regularized by Gaussian function which reflects the harmonic trap. The collective coordinates (dynamical variables) are: 
1) a pair of vortex-core position vectors
and
2) width of Gaussian function and its canonical-conjugate variable.
Our goal is to
derive from equation~(\ref{eq-GP}) the evolution equation for the above collective
coordinates.  TVF for a pair of interacting vortices
with charges $n_i=\pm1$ and $n_j=\pm1$ is thus given by:
\begin{eqnarray}
\Phi \left(t,{\bf r}\right) & = &  Ne^{-\frac{{\bf r}^2}{4w^2 }} \sqrt{\frac{{\left({\bf r}-{\bf r}_i\right)}^2}{2{\xi
}^2+{\left({\bf r}-{\bf r}_i\right)}^2}}\sqrt{\frac{{\left({\bf r}-{\bf r}_j\right)}^2}{2{\xi }^2+{\left({\bf r}-{\bf r}_j\right)}^2}}\nonumber\\
&&\times {\mathrm e}^{i[{\bf r}^2\beta+n_i{{\rm arctan} \left(\frac{y-y_i}{x-x_i}\right)\ }+n_j{{\rm arctan} \left(\frac{y-y_j}{x-x_j}\right)\ }]}\nonumber\\
&\equiv &  N f_0 f_if_j {\mathrm e}^{i(\phi_0+\phi_i+\phi_j)},\nonumber\\
\label{eq-trial-function}
\end{eqnarray}
where $f_0\equiv e^{-\frac{{\bf r}^2}{4w^2 }}, f_k\equiv
\sqrt{\frac{{\left({\bf r}-{\bf r}_k\right)}^2}{2{\xi
}^2+{\left({\bf r}-{\bf r}_k\right)}^2}}$ and
$\phi_0\equiv {\bf r}^2\beta, \quad \phi_k\equiv n_k{{\rm arctan}
\left(\frac{y-y_k}{x-x_k}\right)\ }$ are amplitudes and phases, respectively. 
Here ${\bf r}_k=(x_k,y_k)$ with $k=i,j$ are collective coordinates of vortex-core positions,
and $\xi$ is the healing length related to vortex core size, giving a hallmark of the vortex. 
$w$ in $f_0$ and $\beta$ in $\phi_0$ are the width collective coordinate and its canonical-conjugate variable, respectively.
The normalization factor depends on the positions of vortex cores, and is given by
\begin{eqnarray}\label{norm}
N=\frac{1}{\sqrt{2\pi
w^2-4\pi\xi^2(e^{-\frac{r_i^2}{2w^2}}+e^{-\frac{r_j^2}{2w^2}})}}.
\end{eqnarray}
TVF in equation~(\ref{eq-trial-function}) has a well defined angular momentum.
$w$ stands for the width of the confined region, and the rms $\langle r \rangle$ of the effective area of condensates is represented by $w$ as
\begin{eqnarray}\label{rms}
\langle r \rangle \equiv \sqrt{\bar{r^2}}=\sqrt{\int (x^2+y^2)|\Phi|^2 d^2{\bf r}}\sim \sqrt{2} w. 
\end{eqnarray}

There might be several other variants for TVF, all of which have proved to lead to a set of Lagrange equations which are either contradicting or 
degenerate each other\footnote{Inclusion of collective coordinates (momenta) conjugate to locations of the vortex cores in the phase of TVF worked well in \cite{nakam}. This time, however,  it has proved to lead to the degenerate Lagrange equations, because of strong phase correlations between vortices within a single component. Consequently, one cannot expect the nonzero inertia of vortices.}. 
%\cite{comm}.  
In particular, a different form with use of eigenstates (with non-zero angular momenta) for the 2-d harmonic oscillator was effective in the case of small nonlinearity\cite{garc1,garc3}. Under the strong nonlinearity, however, it turns out giving rise to  an unphysical inter-vortices force that grows with inter-vortices distance, and cannot be suitable as TVF in the present work.

The condition
to minimize the energy $E=\int d^2{\bf r}
\left(|\nabla\Phi|^2+V({\bf r})|\Phi|^2+\frac{g}{2}|\Phi|^4\right)$
evaluated in the limit of noninteracting vortices, leads to the optimal values:
\begin{eqnarray}\label{optim}
w_0^2 &&\cong \frac{1}{4}\sqrt{\frac{g}{\pi}}=O(g^{1/2}),\nonumber\\
\xi &&\cong \frac{|n_i|\pi^{1/4}}{\sqrt{2+\gamma}}g^{-1/4}=O(g^{-1/4}),
\end{eqnarray}
where $\gamma(=0.57721)$ is Euler constant.  We shall take these values in our estimation of orders of 
$w$ and $\xi$, although the following dynamical
treatment will reveal that $w$ will be fluctuating around this optimal value.

First of all we note: GPE in equation~(\ref{eq-GP}) can be derived from the
variational principle that minimizes the action obtained from
Lagrangian density $\mathcal{L}$ for field variables,
\begin{eqnarray}
-\mathcal{L} = \frac{i}{2}(\Phi\dot{\Phi}^\ast-\Phi^\ast\dot{\Phi})+|\nabla\Phi|^2 +{\bf r}^2|\Phi|^2+\frac{g}{2}|\Phi|^4.
\label{eq-lagrangian-density}
\end{eqnarray}
In fact, equation~(\ref{eq-GP}) is obtained from Lagrange equation:
\begin{equation}
\frac{\partial}{\partial
t}\frac{\partial\mathcal{L}}{\partial{\dot{\Phi}^\ast}}-\frac{\partial\mathcal{L}}{\partial\Phi^\ast}+\nabla\frac{\partial\mathcal{L}}{\partial\nabla\Phi^\ast}=0.\label{eq-lagrange-equation}
\end{equation}

By substituting TVF in equation~(\ref{eq-trial-function}) into  equation~(\ref{eq-lagrangian-density}), equation~(\ref{eq-lagrangian-density}) becomes
\begin{eqnarray}
-{\mathcal L}/N^2&=&f^2_0f^2_if^2_j %\nonumber\\ &\times&
 \left( \dot{\beta}\frac{\partial \phi_0}{\partial \beta}+ {\dot{x}}_i\frac{\partial \phi_i}{\partial x_i}+
{\dot{y}}_i\frac{\partial \phi_i}{\partial y_i}+{\dot{x}}_j\frac{\partial \phi_j}{\partial x_j}+{\dot{y}}_j\frac{\partial \phi_j}{\partial y_j}\right)\nonumber\\
&+&{\left(\frac{\partial f_0}{\partial x}f_if_j+f_0\frac{\partial f_i}{\partial x}f_j+f_0f_i\frac{\partial f_j}{\partial x}\right)}^2\nonumber\\
&+&\left(\frac{\partial f_0}{\partial y}f_if_j+f_0\frac{\partial f_i}{\partial y}f_j+f_0f_i\frac{\partial f_j}{\partial y}\right)^2\nonumber\\
&+&f^2_0f^2_if^2_j\left(\frac{\partial \phi_0}{\partial x}+\frac{\partial \phi_i}{\partial x}+\frac{\partial \phi_j}{\partial x}\right)^2\nonumber\\
&+&f^2_0f^2_if^2_j\left(\frac{\partial \phi_0}{\partial x}+\frac{\partial \phi_i}{\partial y}+\frac{\partial \phi_j}{\partial y}\right)^2\nonumber\\
&+&\left(x^2+y^2\right)f_0^2f^2_if^2_j+\frac{gN^2}{2}f_0^4f^4_if^4_j.
\label{eq-lagrangian-density-variational}
\end{eqnarray}

Integrating $\mathcal{L}$ over space coordinates ${\bf r}$, we
obtain the effective Lagrangian $L$ for the collective
coordinates: $L\equiv\int\int d^2 {\bf r}\mathcal{L}$.
Total Lagrangian of the system consists of several integrals:
\begin{eqnarray}
-L/N^2&=&-n_i I_1-n_j I_2+(1+\frac{1}{4w^4}+\dot{\beta}+4\beta^2)I_3 +I_4 -I_5\nonumber \\
& & +n_i^2I_{6}+n_j^2I_{7}+2n_in_jI_{8}+\frac{g}{2}N^2I_{9}-4\beta n_iJ_1-4\beta n_jJ_2,
\label{eq-lagrangian-integrals}
\end{eqnarray}
where all integrals are defined and calculated in Taylor expansions up to $O((\frac{\xi}{w})^2)$
in Table 1 in the end of the text.  In equation~(\ref{eq-lagrangian-integrals}), integrals with their leading order higher than $O((\frac{\xi}{w})^2)$ are suppressed.
The integration related to the  charge-dependent inter-vortex interaction ($I_{8}$) in Table 1 is described in detail in \ref{vort-int}.

In the limit of $\xi/w=O(g^{-\frac{1}{2}})\ll 1$, taking only the
leading order (: $O((\xi/w)^0)$) and using an expansion
$\exp(-\frac{r_k^2}{2w^2})\sim
1-\frac{r_k^2}{2w^2}+\frac{r_k^4}{8w^4}$ in each of the integrals,
$L$ in equation~(\ref{eq-lagrangian-integrals}) with $N$ in equation~(\ref{norm}) is expressed by
\begin{eqnarray}\label{reduced Lag}
-L&=&  \sum\limits_{k=i,j} \left\{-\frac{1}{2w^2} n_k\left(1-\frac{r_k^2}{4w^2}+\frac{r_k^4}{24w^4}\right) \left(y_k\dot{x}_k -x_k\dot{y}_k \right)\right.\nonumber\\
&-&\left.\frac{r_k^2}{6w^4}(1-\frac{r_k^2}{4w^2} )+ \frac{1}{w^2}(1-\frac{r_{k}^2}{2w^2}+\frac{r_k^4}{8w^4})n_{k}^2\right\} \nonumber\\
&+&2w^2\left(1+\dot{\beta}+4\beta^2\right)
+\frac{7}{6w^2}+\frac{g }{8\pi w^2} +V_{ij}. 
\end{eqnarray}
Here, $r_k=\sqrt{x_k^2+y_k^2}$ and
\begin{eqnarray}\label{v-inter}
V_{ij}=-\frac{1}{w^2}U(r_{ij},r_G^{ij})
\end{eqnarray}
with
\begin{eqnarray}\label{magni}
U(r_{ij},r_G^{ij})&\equiv&
n_in_j\left(1 - \frac{{\bf r}_i \cdot {\bf r}_j}{2w^2}\right) \left(\gamma+\ln\left(\frac{r_{ij}^2}{2w^2}\right)
-\frac{r_i^2+r_j^2-{\bf r}_i \cdot {\bf r}_j}{2w^2}\right)\nonumber\\
&+&\frac{5\pi^{1/2}}{2^{5/2}\sqrt{7}r_{ij}^4}e^{-\frac{r_i^2+r_j^2}{4w^2}}I_0\left(\frac{r_G^{ij}r_{ij}}{4w^2}\right).
\end{eqnarray}

On r.h.s. of  equation~(\ref{magni}), the 1st term stands for the
charge-dependent long-range interaction from $I_{8}$ and the 2nd one for the
charge-independent short-range repulsive interaction from $I_4$.
$r_{ij}=\sqrt{x_{ij}^2+y_{ij}^2}$ and
$r_G^{ij}=\sqrt{(x_G^{ij})^2+(y_G^{ij})^2}$ are the inter-vortices
distance and the distance of the center of masses from the origin,
respectively. Other variables used in the calculations are
illustrated in figure~\ref{fig:angle2} in \ref{vort-int}.

Lagrange equations of motion for ${\bf r}_k$ with $k=i,j$ are:
\begin{equation}\label{eq-position}
\frac{d}{dt}\Bigg(\frac{\partial L}{\partial{\dot{{\bf
r}}_k}}\Bigg)-\frac{\partial L}{\partial {\bf r}_k}=0,
\end{equation}
and those for $\beta$ and $w$ are:
\begin{eqnarray}\label{eq-width}
\frac{d}{dt}\Bigg(\frac{\partial L}{\partial{\dot{\beta}}}\Bigg)-\frac{\partial L}{\partial \beta}=0,\nonumber\\
\frac{d}{dt}\Bigg(\frac{\partial L}{\partial{\dot{w}}}\Bigg)-\frac{\partial L}{\partial w}=0.
\end{eqnarray}

In the asymptotic region ($\xi \ll r_{ij}, r_G^{ij}  \ll w$) with
$g\gg1$, these equations reduce to:
\begin{eqnarray}\label{coup-eq-cores}
n_k \dot{y_k}&=&(\frac{1}{3}+n_{k}^2)(1+\frac{r_k^2}{2w^2})\frac{x_k}{w^2}+ n_k \frac{\dot{w}}{w} y_k  
-  w^2(1+\frac{r_k^2}{2w^2})\frac{\partial V}{\partial x_k}, \nonumber\\
n_k
\dot{x_k}&=&-(\frac{1}{3}+n_{k}^2)(1+\frac{r_k^2}{2w^2})\frac{y_k}{w^2}+
n_k\frac{\dot{w}}{w} x_k
+w^2(1+\frac{r_k^2}{2w^2})\frac{\partial V}{\partial y_k},
\end{eqnarray}
and
\begin{eqnarray}\label{coup-eq-wd}
\ddot{w}&=&-4w + \frac{g}{4\pi w^3} 
 - \frac{1}{w^5}(\sum_{k=i,j}(\frac{1}{3}+n_{k}^2)r_k^2)
-\frac{\partial V}{\partial w}, \nonumber\\
\beta&=&\frac{\dot{w}}{4w},
\end{eqnarray}
where $V\equiv V_{ij}$.
Equations (\ref{coup-eq-cores}) and (\ref{coup-eq-wd}) together with equations (\ref{v-inter}) and (\ref{magni})
shows:

1) Effective dynamics of interacting vortices is coupled with time-derivative of width $w$.
As we shall see in the numerical analysis of the effective dynamics and GPE in the next section, the second term on r.h.s of equation~(\ref{coup-eq-cores})  plays an important role to guarantee the width-induced breathing oscillations (or radial oscillations ) of each vortex and thereby the validity of the variational approach in the present paper;

2) As seen in equations (\ref{v-inter}) and (\ref{magni}), the inter-vortex interaction consists of
a short-range repulsive interaction which is independent from vortex charges and a charge-dependent long-range logarithmic interaction multiplied with the pre-exponential factor which depends on the scalar product of a pair of position vectors of vortex cores;

3) As seen in the first term on r.h.s. of equation~(\ref{coup-eq-cores}), the nonlinear spring force has a
charge-independent contribution $\propto \frac{1}{3}$ which is additive to 
the charge-dependent one $\propto n^2_k$.

4) Equation for $w$ in equation~(\ref{coup-eq-wd}), if its coupling with vortex cores would be suppressed,
can lead to a linearized equation as
\begin{eqnarray}\label{breathing}
\ddot{w}=-16(w-w_0),
\end{eqnarray}
with use of the optimal value $w_0$ in equation~(\ref{optim}). The angular frequency($\omega_0=\sqrt{16}=4$) or frequency($\nu=\frac{4}{2\pi}$) here accords with the those of the breathing mode\cite{Stri,Jin,Mew,fett1,fett2}. More general evaluation of the breathing mode is described in \ref{brm-no vortex}.

5) If we shall freeze the width degree of freedom (: $\dot{w}=0, w=w_0$),
the effective dynamics in equation~(\ref{coup-eq-cores}) is similar to the one employed in the preceding works in Refs. 
\cite{torr,nava,Kou,Kol} which has also a nonlinear spring force and a renormalization of the strength of the inter-vortex interaction but has no coupling with the width degree of freedom. The minor discrepancy of the spring force constant and renormalized strength of inter-vortex interaction between the present and preceding works are due to the Thomas-Fermi approximation in the latter works which suppresses the kinetic energy term in GPE.

While equation~(\ref{coup-eq-cores}) is essential,  its more convenient form is
available by using scaled coordinates,
\begin{eqnarray}\label{scale-cor}
X_k\equiv\frac{x_k}{w}, \quad
Y_k\equiv\frac{y_k}{w}.
\end{eqnarray}
In fact, $X_k$ and $Y_k$ prove to satisfy:
\begin{eqnarray}\label{coup-eq-scv}
n_k \dot{Y_k}&=&\frac{1}{w^2}(\frac{1}{3}+n_{k}^2)(1+\frac{X_k^2+Y_k^2}{2})X_k 
- (1+\frac{X_k^2+Y_k^2}{2})\frac{\partial
V}{\partial X_k},
\nonumber\\
n_k \dot{X_k}&=&-\frac{1}{w^2}(\frac{1}{3}+n_{k}^2)(1+\frac{X_k^2+Y_k^2}{2})Y_k
+(1+\frac{X_k^2+Y_k^2}{2})\frac{\partial V}{\partial Y_k},
\end{eqnarray}
which have hided the second terms on r.h.s of equation~(\ref{coup-eq-cores}).  

By introducing $R_k^2\equiv X_k^2+Y_k^2$ and the nonstandard Poisson bracket defined by
\begin{eqnarray}\label{poiss}
\fl \{A,B\} \equiv \frac{\partial A}{\partial p_w}\frac{\partial B}{\partial w}
-\frac{\partial A}{\partial w}\frac{\partial B}{\partial p_w} 
+
\sum_{k=i,j}(\frac{\partial A}{\partial Y_k}\frac{\partial B}{\partial X_k}
-\frac{\partial A}{\partial X_k}\frac{\partial B}{\partial Y_k})\cdot(1+\frac{R_k^2}{2}), 
\end{eqnarray}
one can construct the Hamiltonian
formalism corresponding to equations (\ref{coup-eq-scv}) and (\ref{coup-eq-wd}) as follows:
\begin{eqnarray}\label{inert1 eq}
n_k \dot{X_k}&=&\{H,X_k\}=-(1+\frac{R_k^2}{2})\frac{\partial H}{\partial Y_k},
\nonumber\\
n_k \dot{Y_k}&=&\{H,Y_k\}=(1+\frac{R_k^2}{2})\frac{\partial H}{\partial X_k},\nonumber\\
\dot{w}&=&\{H,w\}=\frac{\partial H}{\partial p_w}, \quad\dot{p}_w=\{H,p_w\}
=-\frac{\partial H}{\partial w}, 
\end{eqnarray}
with Hamiltonian $H$  given by
\begin{eqnarray}\label{inert2 eq}
\fl H=(\frac{p_w^2}{2}+2w^2+\frac{g}{8\pi w^2})  
+ \sum\limits_{k=i, j}\frac{1}{2w^2}(\frac{1}{3}+n_{k}^2)R_k^2+\frac{1}{w^2}U(wR_{ij},w\Lambda_G^{ij}). 
\end{eqnarray}
In equation~(\ref{inert2 eq}), by using  equations (\ref{v-inter}) and (\ref{magni}), $-V_{ij}$ has been replaced by $\frac{1}{w^2}U(wR_{ij},w\Lambda_G^{ij})$ with the scaled variables $R_{ij}\equiv\frac{r_{ij}}{w}$ and 
$\Lambda_G^{ij}\equiv \frac{r_G^{ij}}{w}$.

Now we shall proceed to the numerical analysis of 
both the effective dynamics and GPE.

\section{Numerical analysis}\label{sec-some-vortices}
\subsection{Numerical test of the effective dynamics}

\begin{figure}[!htbp]
\centering
\includegraphics[width=0.65\linewidth]{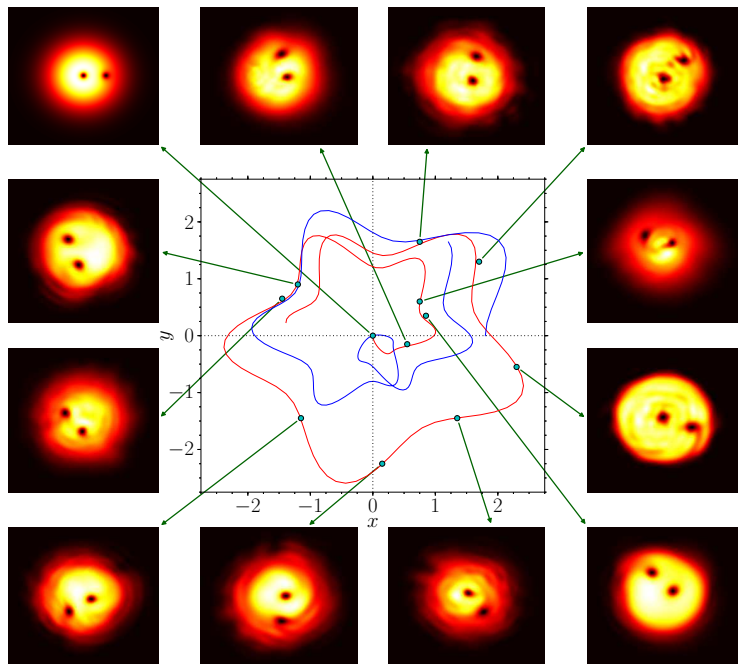}
\caption{(Color online) Short-time trajectories available from the
Hamilton dynamics described by equations (\ref{inert1 eq}) with
(\ref{inert2 eq}) and snapshots of the wave function amplitudes obtained from GPE in
equation~(\ref{eq-GP}) at several distinct time in the case of two vortices  with
identical charges $(n_i=n_j=1)$. The scale of time is common to both 
Hamilton dynamics and GPE.
$g=500$. Vortex cores $i$ (blue line) and $j$(red line) start from initial values $(x_i(0), y_i(0))=(1.8, 0)$ 
and $(x_j(0), y_j(0))=(0, 0)$, respectively. Other initial values are 
$w(0)(=(1/2)(g/\pi)^{1/4})=1.78$ (see equation~(\ref{optim})) and
$p_{w}(0)=0$. Both cores show counter-clockwise motions. Each snapshot of wave function is assigned to the trajectory of the core  $j$.} 
\label{snap}
\end{figure}

With use of $g=500$, we shall show in figure~\ref{snap} how the effective dynamics described by 
equations (\ref{poiss})-(\ref{inert2 eq}) is firmly supported by the numerical wave dynamics based on GPE in equation~(\ref{eq-GP})(as for algorithms involving real and imaginary-time propagation based on a split-step Crank-Nicolson method, see, e.g., \cite{Murug2009,Vudra2012} ). 
This is short-time trajectories for a pair of vortices with identical charges ($n_i=n_j=1$). The positions of vortex cores found in the analysis of GPE are nicely reproduced by  the effective dynamics. The agreement of vortex-core positions between the wave dynamics of GPE and the effective 
dynamics is found for initial vortex configurations other than those given figure~\ref{snap}. Here it should be remarked: each vortex core shows the rapid oscillation in radial direction, besides its slower rotational motion.  This radial oscillation is induced by breathing of the width, and can not be produced in the effective dynamics under the fixed value of width $w$.

\begin{figure}[htb]
\centering
\includegraphics[width=0.65\linewidth]{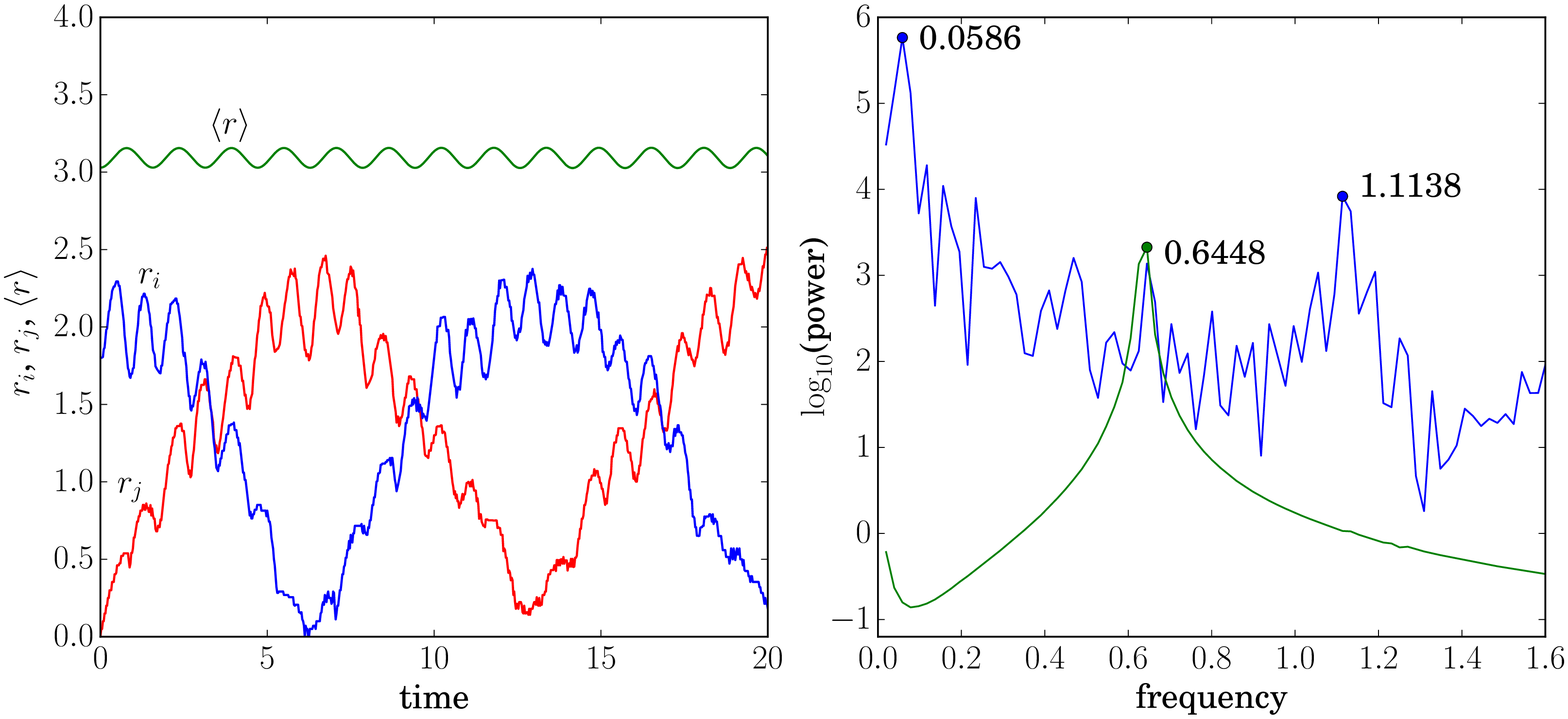}
\includegraphics[width=0.65\linewidth]{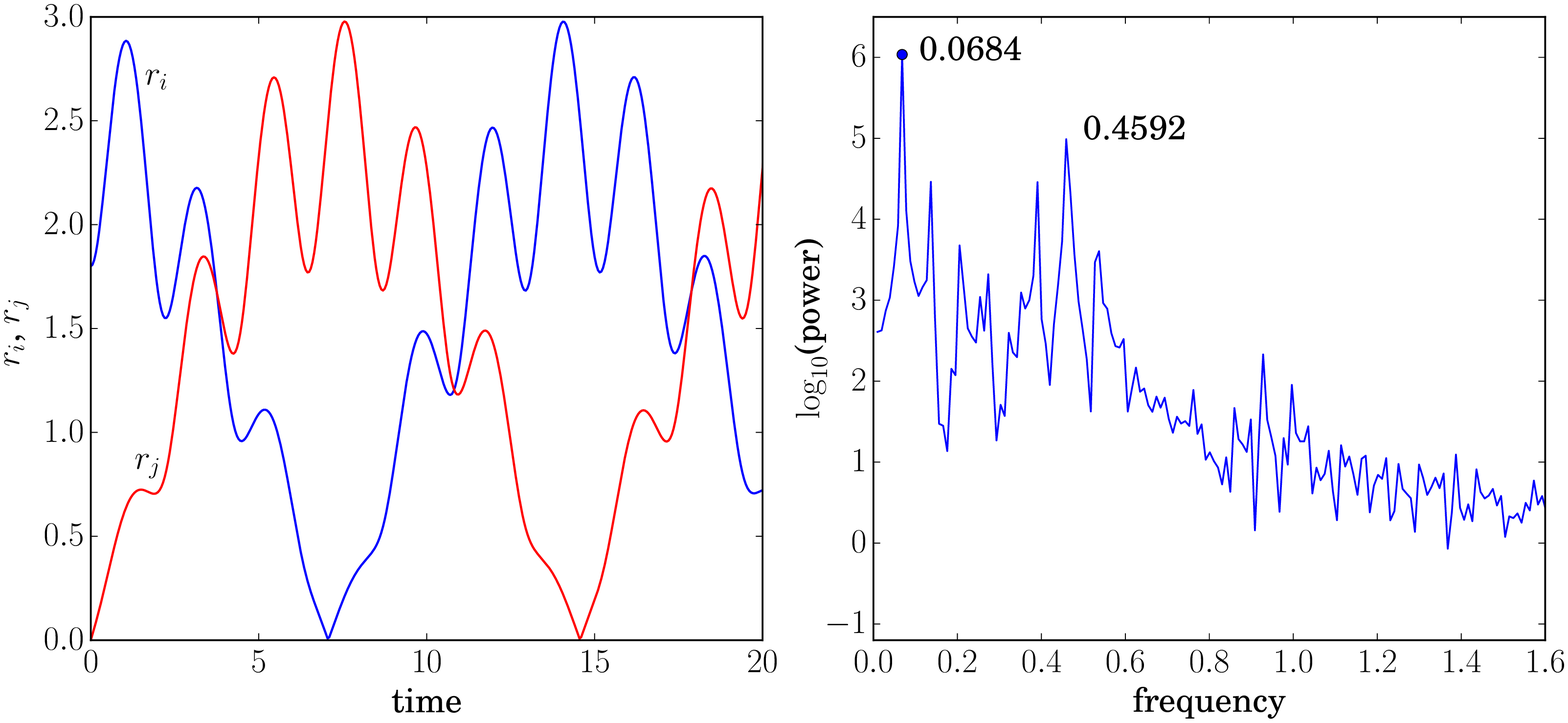}
\includegraphics[width=0.65\linewidth]{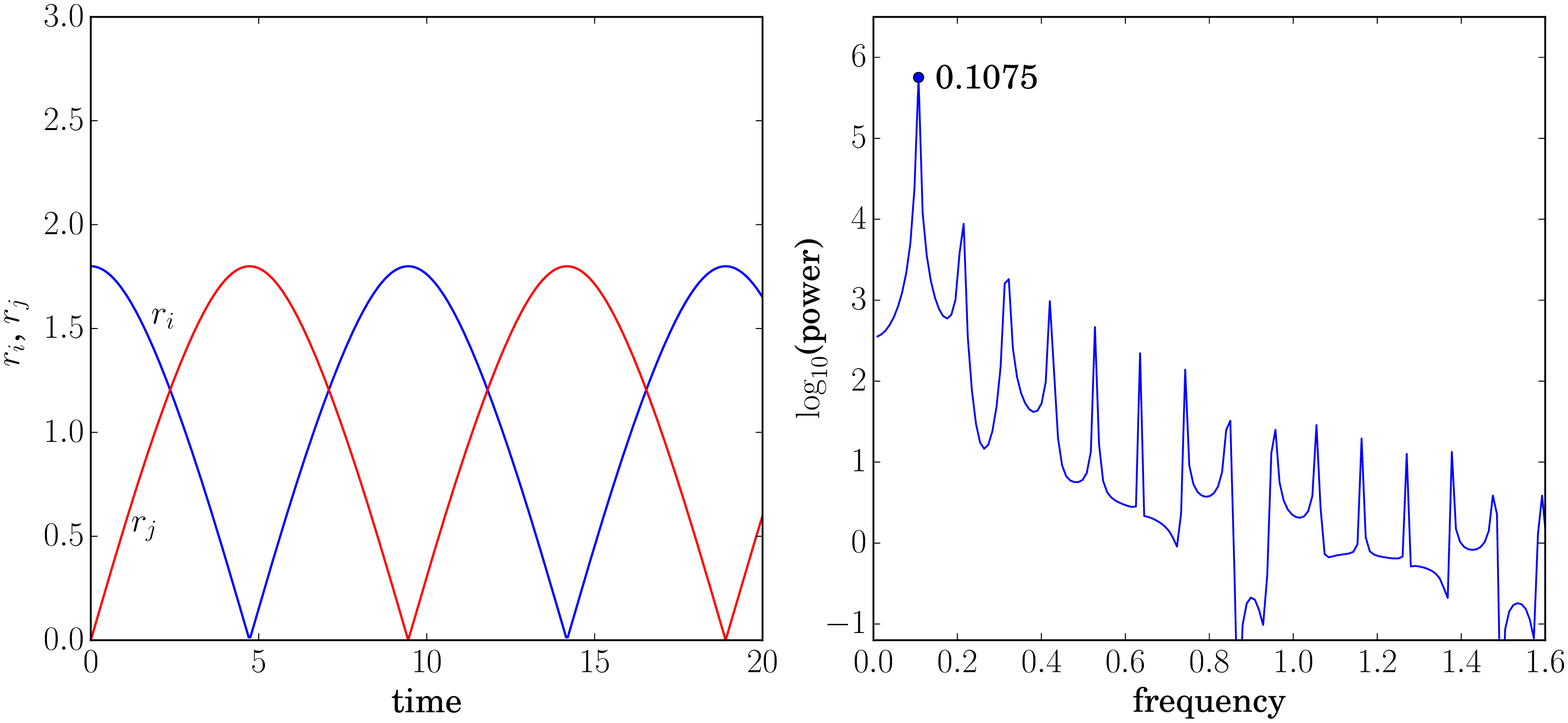}
\caption{(Color online) Left panels: Time ($t$) dependence of  radial distance 
$\bar{r}_i, \bar{r}_j$ of each vortex core. Red and blue are for $\bar{r}_j$ and $\bar{r}_i$, respectively.
Right panels: $\nu$ dependence of the corresponding power spectra, where red and blue are degenerate.
Top, middle and bottom panels are results computed by wave dynamics of GPE in equation~(\ref{eq-GP}),
effective dynamics in 
equations (\ref{poiss})-(\ref{inert2 eq}), and effective dynamics with frozen width degree of freedom, respectively.
Top panel includes rms $ \langle r \rangle$ (blue) of the effective area of condensates defined by equation~(\ref{rms}).
$g=500$ and initial values are the same as in figure~\ref{snap}. } 
\label{freqs}
\end{figure}

Figure \ref{freqs} shows the time ($t$) dependence of the radial distance $\bar{r}_i, \bar{r}_j$ of each vortex core and its power spectra. They are computed in three ways: [i]wave dynamics of GPE in equation~(\ref{eq-GP}); [ii]effective dynamics including  width degree of freedom in equations (\ref{poiss})-(\ref{inert2 eq}); [iii]effective dynamics with frozen width degree of freedom (i.e., under the fixed value of width in equation~(\ref{optim})). The case [i] includes rms $ \langle r \rangle$ (green) of the effective area of condensates defined by equation~(\ref{rms}).

From the left-top panel, we see that wave dynamics of GPE shows the slow circular oscillation of vortex cores around the origin  superposed on the rapid radial oscillation (: breathing mode). This characteristic is reproduced by the effective dynamics including the width degree of freedom (left-middle panel), but not by the effective dynamics with frozen width degree of freedom (left-bottom panel).  From the power spectra on the right-top panel,
we see that wave dynamics of GPE leads to three peaks at $\nu_1=0.0586, \nu_2=0.6448$ and $\nu_3=1.1138$, which represent
the slow circular motion, breathing mode and minor fluctuations of core width, respectively. 
The frequency
$\nu_2$, which is also the frequency of  $\langle r \rangle$, is attributed to the free breathing mode with its frequency $\nu=\frac{\omega_0}{2\pi}
=\frac{4}{2\pi}=0.636$ available from equation~(\ref{breathing}) in the effective dynamics.

In the effective dynamics with width degree of freedom (right-middle panel) the lower two peaks  $\nu_1'=0.0684$ and $\nu_2'=0.4592$  correspond to circular motion and breathing modes, respectively. But we see no counterpart of fluctuations of core width involved in the wave dynamics, because such fluctuations are not taken into consideration in the present variational principle. Small discrepancy of the lower two peaks ($\delta \nu_1=\nu_1'-\nu_1, \delta \nu_2=\nu_2'-\nu_2$) between the wave dynamics and effective dynamics is found to decrease as the value of $g$ is further increased. 

In the effective dynamics with frozen width degree of freedom (right-bottom panel), we see neither the correct peak for the circular motion nor the breathing mode.

\begin{figure}[htb]
\centering
\includegraphics[width=0.45\linewidth]{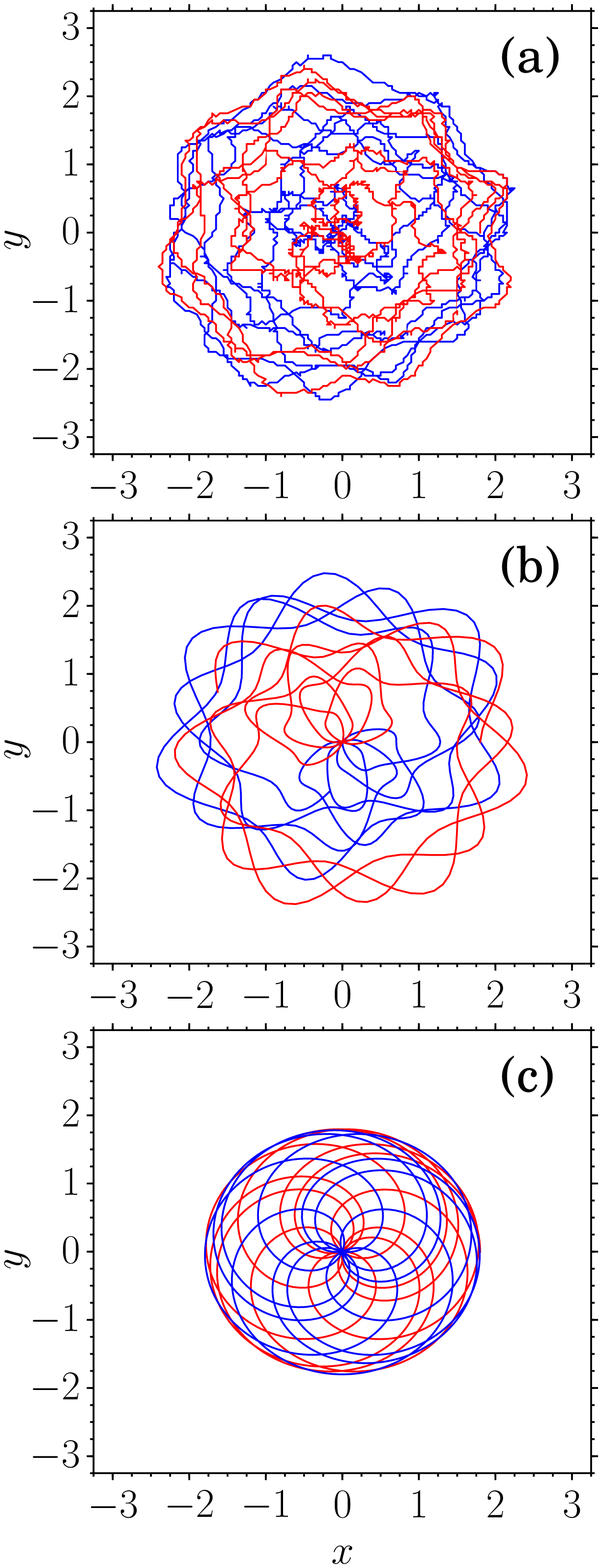}
\caption{(Color online) Trajectories of
two vortex cores $i$(red) and $j$(blue)  with identical vortex charges
$(n_i=n_j=1)$. $g=800$. Initial values: $x_i(0)=1.8,\quad y_i(0)=0,\quad x_j(0)=0,\quad y_j(0)=0,\quad w(0)=1.7, \quad p_w(0)=0.1$.  (a) Result obtained by the
iteration of GPE in equation~(\ref{eq-GP}); (b) Result obtained by
iterating the Hamilton equation in equations (\ref{inert1 eq}) with
(\ref{inert2 eq}) ; (c) Result obtained by
iterating the Hamilton equation in equations (\ref{inert1 eq}) with
(\ref{inert2 eq}) with frozen width degree of freedom.
} \label{two-same}
\end{figure}

\begin{figure}[htb]
\centering
\includegraphics[width=0.45\linewidth]{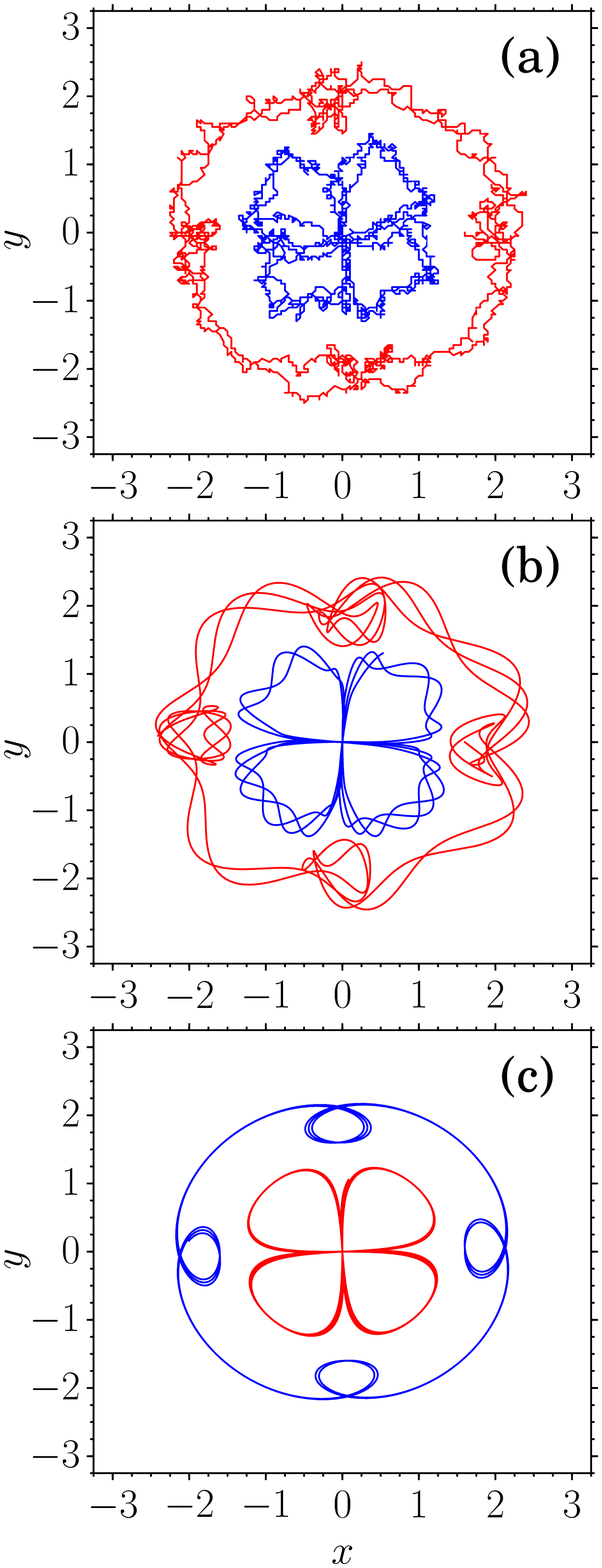}
\caption{(Color online) Dipole dynamics, i.e., trajectories of two vortices dynamics with different vortex charges$(n_i=1, n_j=-1)$. $g=800$. Initial values: $x_1(0)=1.6,\quad y_1(0)=0,\quad x_2(0)=0,\quad y_2(0)=0,\quad w(0)=2, \quad p_w(0)=1$. The meaning of subfigures (a)-(c) is the same as in figure~\ref{two-same}. }
\label{two-oppose}
\end{figure}

\subsection{Geometry of trajectories of a pair of vortex cores}
We shall proceed to show the geometry of longer-time trajectories of vortex cores  in case of $g=800$.
First we investigate a pair of vortex cores  with the same charges $n_i=n_j=1$.
Figure \ref{two-same} shows the results of (a)wave dynamics of GPE in equation~(\ref{eq-GP}), (b) 
Hamilton equation including the width degree of freedom in equations (\ref{inert1 eq}) and
(\ref{inert2 eq}), and (c)Hamilton equation with frozen width degree of freedom. 
We find that the trajectories  generated by the effective dynamics
well mimic the corresponding ones obtained by wave dynamics of GPE. This fact justifies the validity of both TVF in equation~(\ref{eq-trial-function}) and  effective dynamics 
in equations (\ref{inert1 eq}) with
(\ref{inert2 eq}). The effective dynamics with frozen width degree of freedom can reproduce neither rippled structures (:  rapid radial oscillation superposed on the slower rotational motion) nor the global geometry of trajectories. Likewise, the effective dynamics proposed so far\cite{torr,nava,Kou,Kol} does not include the coupling with $w(t)$, and therefore cannot generate these rippled structures.

On the other hand, two vortices with the different charges $n_i=1,n_j=-1$ show a dipole dynamics.
Figure \ref{two-oppose} again shows  a nice agreement of the results between the  wave dynamics based on GPE and the effective dynamics including the width degree of freedom. The effective dynamics with frozen width degree of freedom cannot generate the rippled structure exhibited by the wave dynamics.

\subsection{Structure of sea-urchin needles and chaos in higher energies}

Encouraged by the effectiveness of the variational approach, we shall now analyze  the dynamics of two vortices in higher energies.
As is recognized in the canonical scheme in equations (\ref{poiss})-(\ref{inert2 eq}),
the trapped two-vortex system has 3 coupled degrees of freedom (: 2 degrees for two vortex cores 
nd 1 degree for width) but possesses  2 constants of motion (total energy and $z$ component of angular momentum). 
Then Poincar\'e-Bendixon's theorem guarantees the nonintegrability and
chaos. This conjecture is different from the consensus of hydrodynamics of interacting point vortices in 2 dimensions, where the number of vortices must be larger than 3 to make the system chaotic\cite{rf:aref}.

\begin{figure}
\centering
\includegraphics[width=0.33\linewidth]{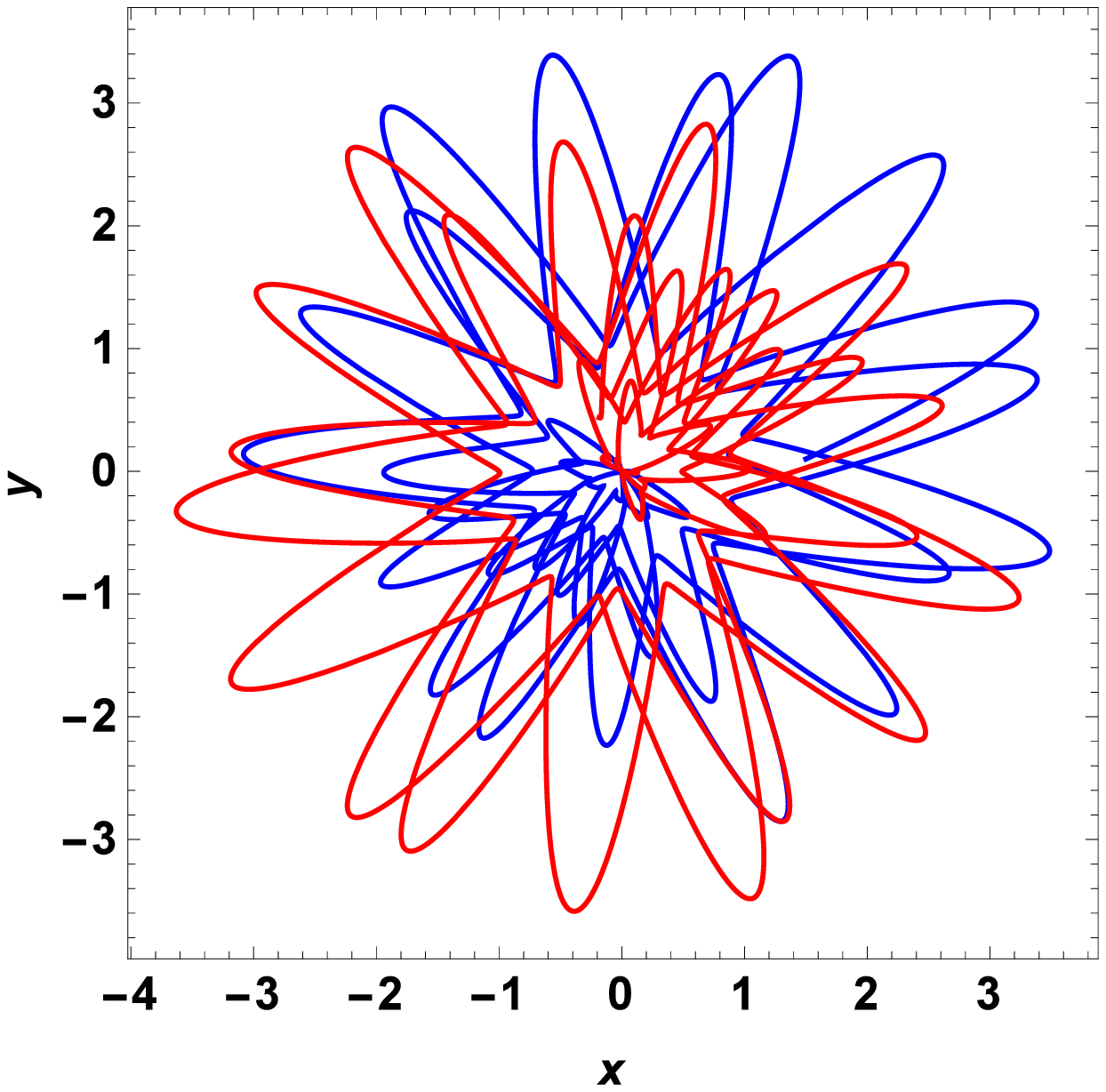}
\includegraphics[width=0.33\linewidth]{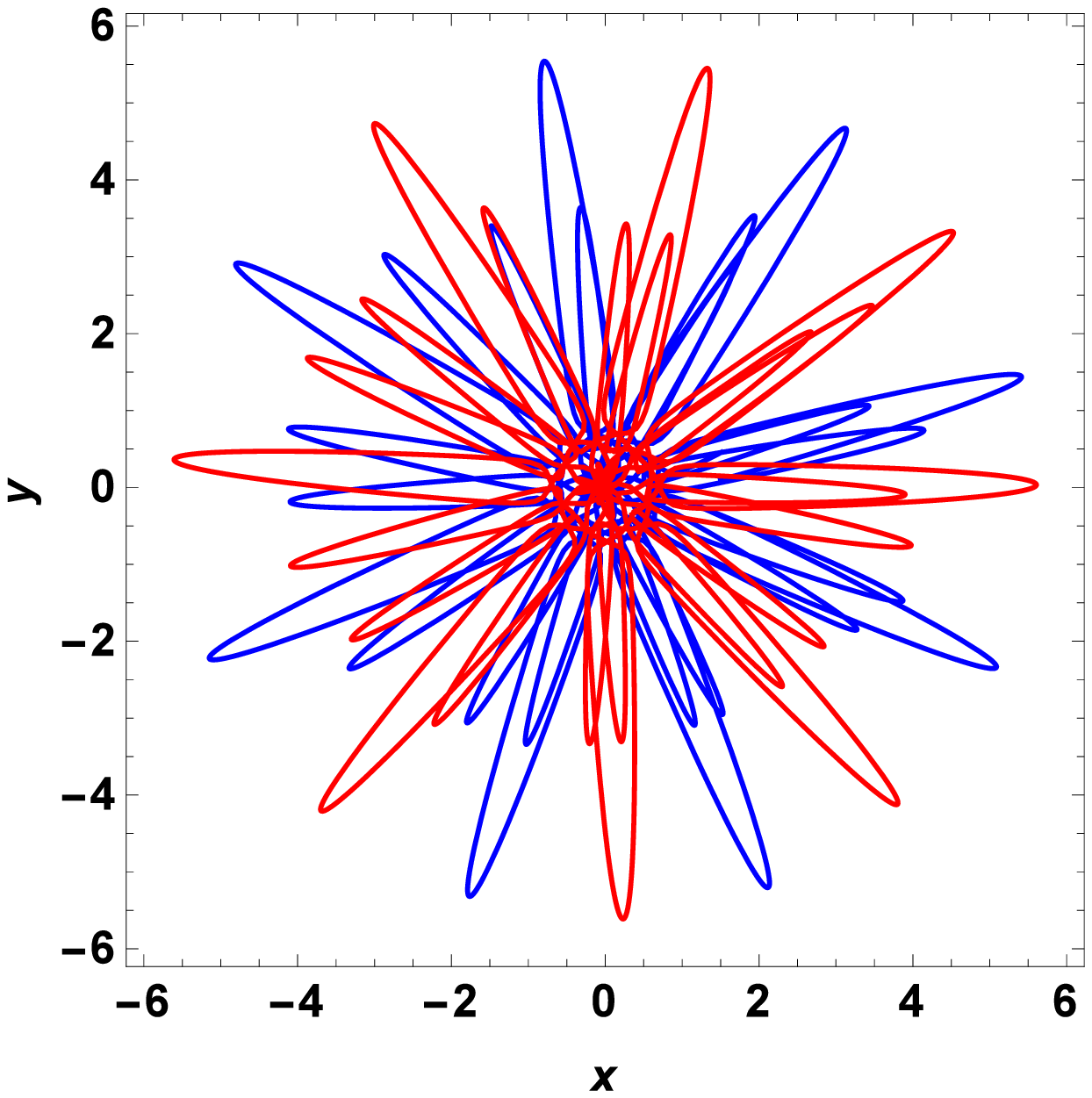}
\includegraphics[width=0.33\linewidth]{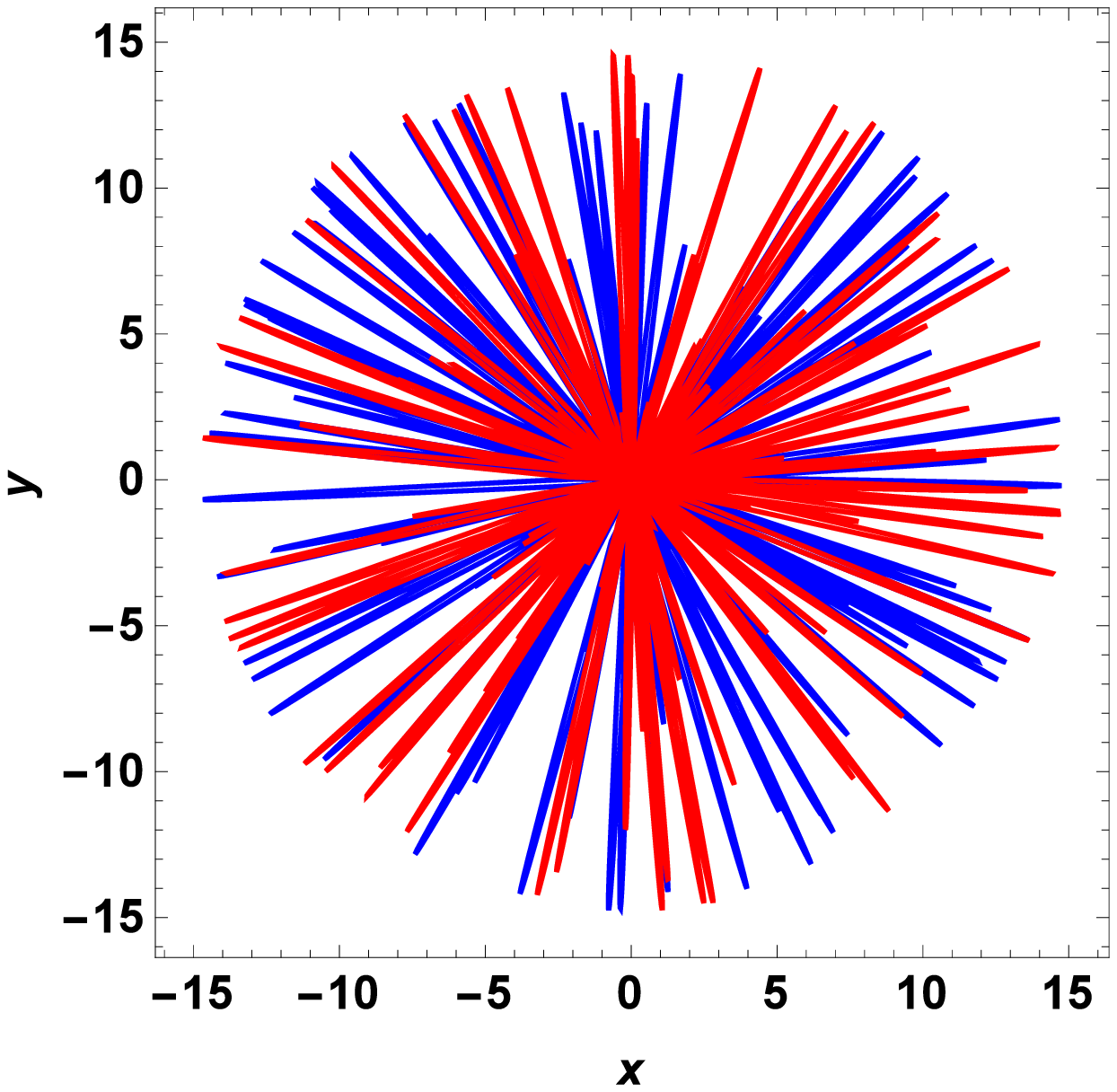}
\includegraphics[width=0.33\linewidth]{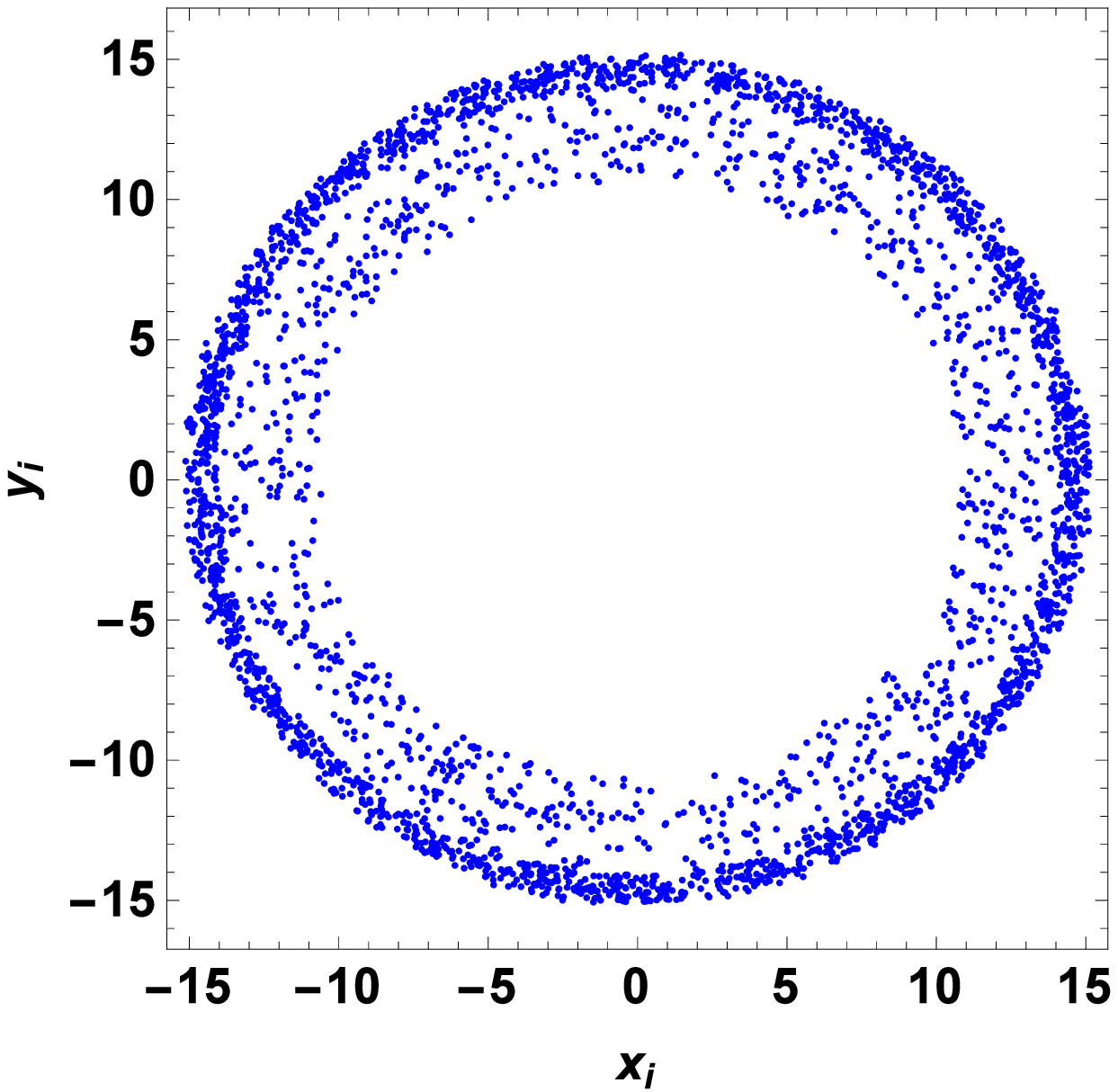}
\caption{(Color online) Trajectories  and Poincar\'e cross section of two vortices with identical charges
($n_i=n_j=1$) in higher energies.  
$g=800; x_i(0)=1.5; y_i(0)=0.1; x_j(0)=0; y_j(0)=0; w(0)=1.55$.
Upper-left panel: trajectories for $p_w(0)=5$; Upper-right panel: trajectories for $p_w(0)=10$; Lower-left panel: trajectories for $p_w(0)=30$; 
Lower-right panel: Poincar\'e cross section $p_w(0)=30$.}
\label{regular}
\end{figure}

Below we shall confine to the case of the identical charges and show how the long-time trajectories of vortex cores will vary by increasing
the initial momentum $p_w(0)$ with other initial variables kept unchanged.
Figure \ref{regular}  shows that both the amplitude  and frequency of the breathing-induced radial oscillation become larger as the system's extra energy (proportional to the square of $p_w(0)$ ) is increased. 
The rippled circular structure in low energy
 ($p_w(0)=5$) changes to a structure of sea-urchin needles in high energy ($p_w(0)=30$). 
This is another interesting issue of breathing dynamics.

In the case of $p_w(0)=30$, we also show  Poincar\'e cross section (see, lower-right panel) composed at each time that 
$w(t)$ has temporally-local maxima, which clearly indicates emergence of chaos in two vortex systems. 
On the other hand, we find that the increase of $p_w(0)$ leads to no chaos in the width dynamics, but only to the increase of the amplitude of its periodic  oscillation. Therefore the emergence of chaos in two vortex systems is caused by the breathing oscillation of $w$ which plays a role of the periodic kicking in the kicked rotator.

The above results hold for vortices with the identical charges.
In the actual numerical analysis, the charge-independent  repulsive potential 
present in the last term of  equation~(\ref{magni})
plays an important role. Vortices with identical charges show commonly clock-wise (or counter-clock-wise) motions and have little chance to become degenerate when this repulsive divergent force become operative. Therefore the numerical analysis of long-time dynamics is possible in the case of identical charges. By contrast, vortices with different charges show counter-rotating motions each other and often  become degenerate, when this divergent force requires a very high precision of computation for us to obtain reliable long-time data. 
On the other hand, the effective dynamics here is concerned with the Hamiltonian system with 3 degrees of freedom. Analysis of its phase-space structure, onset of chaos, Lyapunov exponent, etc is interesting, but requires more extensive study. All these remaining numerical problems will be analyzed elsewhere.

\section{Conclusion}\label{sec-conclusion}
We investigated a role of breathing width degree of freedom  in the effective theory of interacting vortices in a trapped single-component Bose-Einstein condensates (BECs) in 2
dimensions under the strong repulsive cubic nonlinearity.
We have obtained 
Lagrange equation for the interacting vortex-core coordinates coupled with the time-derivative of width and constructed its Hamilton formalism by having recourse to a non-standard Poisson bracket. 
By solving the Hamilton equation, we find
the rapid radial breathing oscillations superposed on the slower rotational motion of vortex cores,  justified by numerical solutions of Gross-Pitaevskii equation. In other words, the rapid radial fluctuations observed in the wave dynamics cannot be explained within the  frame of effective dynamics with a frozen width degree of freedom.

In 2 vortex systems with the identical charges in higher energies,  the breathing oscillation of width plays role of a kicking in the kicked rotator. As the kicking strength increases, both the amplitude and frequency of the breathing-induced radial oscillation become large and the trajectory of each vortex forms a structure of sea-urchin needles.
Byproduct of the present variational approach includes  the logarithmic inter-vortex interaction multiplied with a pre-factor which depends on the scalar product of a pair of core-position vectors, a short-range repulsive inter-vortex interaction  and a charge-independent spring force additive to the charge-dependent one.

%{\em Acknowledgments.} 
\ack One of the authors (K. N.) is grateful to F. Abdullaev, F. Mertens, 
Y. Musakhanov, V.M. P\'erez-Garc\'{\i}a and M. Tsubota for enlightening discussions in the early stage of this work. The work of P. M. forms a part of Science \& Engineering Research Board, Department of Science \& Technology, Govt. of India sponsored research project (SERB Ref. No. EMR/2014/000644).

%\newpage
%\begin{widetext}\label{tab-int}

\begin{table}[!ht]
\caption{Expressions for all integrals in equation (\ref{eq-lagrangian-integrals}), expanded up to $O((\frac{\xi}{w})^2)$.}
\begin{center}
\begin{tabular}{|p{0.4in}|p{2.6in}|p{2.5in}} \hline
Integ- rals & Definitions & Integration results
\\ \hline $I_1$ & $\int d^2{\bf r} e^{-\frac{{\bf r}^2}{2w^2}}\frac{(x-x_i)\dot{y_i}-(y-y_i)\dot{x_i}}{2\xi^2+({\bf r}-{\bf r}_i)^2}\frac{({\bf r}-{\bf r}_j)^2}{2\xi^2+({\bf r}-{\bf r}_j)^2} $ & $2\pi(y_i\dot{x_i}-x_i\dot{y_i})\left(\frac{w^2}{r_i^2}(1-e^{-\frac{r_i^2}{2w^2}})-\frac{\xi^2}{w^2}e^{-\frac{r_i^2}{2w^2}}\right)+\frac{4\sqrt{\pi}\xi^2}{r_{ij}^2}[(x_i-x_j)\dot{y_i}-(y_i-y_j)\dot{x_i}]\times e^{-\frac{r_i^2+r_j^2}{4w^2}}I_0(r_{ij}r_G^{ij}/{2w^2}) $
\\ \hline $I_2$ & $\quad i \leftrightarrow j \quad {\rm in}\quad I_1$ & $\quad i \leftrightarrow j \quad {\rm in}\quad I_1 $
\\ \hline $I_3$ & $\int d^2{\bf r} e^{-\frac{{\bf r}^2}{2w^2}}{\bf r}^2\frac{({\bf r}-{\bf r}_i)^2}{2\xi^2+({\bf r}-{\bf r}_i)^2}\frac{({\bf r}-{\bf r}_j)^2}{2\xi^2+({\bf r}-{\bf r}_j)^2} $ & $4\pi w^4+8\pi w^2\xi^2-4\pi\xi^2\Big[(2w^2+ r_i^2)e^{-\frac{r_i^2}{2w^2}}+(i \leftrightarrow j)\Big]$
\\ \hline $I_4$ & $4\xi^4\int d^2{\bf r} e^{-\frac{{\bf r}^2}{2w^2}}\frac{1}{[2\xi^2+({\bf r}-{\bf r}_i)^2]^3}\frac{({\bf r}-{\bf r}_j)^2}{2\xi^2+({\bf r}-{\bf r}_j)^2} + (i \leftrightarrow j )$ & $\frac{2\pi}{3}\left(e^{-\frac{r_i^2}{2w^2}}(1+\frac{\xi^2r_i^2}{3w^4})+(i \leftrightarrow j)\right)-\left(\frac{5\pi^{3/2}}{2^{3/2}\sqrt{7}r_{ij}^4}-\frac{6\pi^{3/2}\xi^2}{\sqrt{5}r_{ij}^6}\right)e^{-\frac{r_i^2+r_j^2}{4w^2}}I_0\left(\frac{r_G^{ij}r_{ij}}{4w^2}\right) $
\\ \hline $I_5$ & $\frac{2\xi^2}{w^2}\int d^2{\bf r} e^{-\frac{{\bf r}^2}{2w^2}}\frac{x(x-x_j)+y(y-y_j)}{[2\xi^2+({\bf r}-{\bf r}_j)^2]^2}\frac{({\bf r}-{\bf r}_i)^2}{2\xi^2+({\bf r}-{\bf r}_i)^2} + (i \leftrightarrow j ) $ & $\frac{\pi\xi^2}{32w^2}\Bigg[ e^{-\frac{r_i^2}{2w^2}}\left(1-\frac{2^5r_i^2}{w^2}\right)+(i\leftrightarrow j)\Bigg]$
\\ \hline $I_{6}$ & $\int d^2{\bf r} e^{-\frac{{\bf r}^2}{2w^2}}\frac{({\bf r}-{\bf r}_i)^2}{[2\xi^2+({\bf r}-{\bf r}_i)^2][2\xi^2+({\bf r}-{\bf r}_j)^2]} $ & $2\pi e^{-\frac{r_i^2}{2w^2}} -4\pi\xi^2\Bigg[\frac{1}{w^2}e^{-\frac{r_i^2}{2w^2}}
( 1-\frac{r_i^2}{2w^2})
+\frac{2\sqrt{\pi}}{r_{ij}^2}e^{-\frac{r_i^2+r_j^2}{4w^2}}
I_0(\frac{r_{ij}r_G^{ij}}{2w^2})\Bigg] $
\\ \hline $I_{7}$ & $\quad i \leftrightarrow j \quad {\rm in}\quad I_{6}$ & $\quad i \leftrightarrow j \quad {\rm in}\quad I_{6} $
\\ \hline $I_{8}$ & $\int d^2{\bf r} e^{-\frac{{\bf r}^2}{2w^2}}\frac{({\bf r}-{\bf r}_i)\cdot ({\bf r}-{\bf r}_j)}{2\xi^2+({\bf r}-{\bf r}_i)^2}\frac{1}{2\xi^2+({\bf r}-{\bf r}_j)^2} $ & $-\pi e^{-\frac{l_G^2}{2w^2}+\frac{r_{ij}^2}{8w^2}}\left(\gamma+\ln\left(\frac{r_{ij}^2}{2w^2}\right)- \frac{4(r_G^{ij})^2+3r_{ij}^2}{8w^2}\right) $
\\ \hline $I_{9}$ & $\int d^2{\bf r} e^{-\frac{{\bf r}^2}{w^2}}\left(\frac{({\bf r}-{\bf r}_i)^2}{2\xi^2+({\bf r}-{\bf r}_i)^2}\frac{({\bf r}-{\bf r}_j)^2}{2\xi^2+({\bf r}-{\bf r}_j)^2}\right)^2$ & $\pi w^2-8\pi\xi^2\left(e^{-\frac{r_i^2}{w^2}}+e^{-\frac{r_j^2}{w^2}}\right)$
\\ \hline $J_1$ & $\int d^2{\bf r} e^{-\frac{{\bf r}^2}{2w^2}}\frac{x(y-y_i)+y(x-x_i)}{2\xi^2+({\bf r}-{\bf r}_i)^2}\frac{({\bf r}-{\bf r}_j)^2}{2\xi^2+({\bf r}-{\bf r}_j)^2}$ & $ -4\xi^2\pi^{3/2}
\frac{x_Gy_{ij}-y_Gx_{ij}}{r_{ij}^2} e^{-\frac{r_i^2+r_j^2}{4w^2}}I_0\left(\frac{r_{12}r_G^{ij}}{4w^2}\right)$
\\ \hline $J_2$ &$\quad i \leftrightarrow j \quad {\rm in}\quad J_1$ & $\quad i \leftrightarrow j \quad {\rm in}\quad J_1$ \\
\hline
\end{tabular}
\end{center}
\end{table}
%\end{widetext}

\appendix
\section{Calculation of the charge-dependent inter-vortex interaction and $I_{8}$}\label{vort-int}
This interaction is due to the integral
\begin{eqnarray}
V_{tp}&=&
2N^2\int\limits_{-\infty}^{\infty}\int\limits_{-\infty}^{\infty} f_0^2f_i^2 f_j^2\left(\frac{\partial \phi_i}{\partial x}\frac{\partial \phi_j}{\partial x}+\frac{\partial \phi_i}{\partial y}\frac{\partial \phi_j}{\partial y}\right) d^2{\bf r}\nonumber\\
&=&2N^2n_in_j\int\limits_{-\infty}^{\infty}\int\limits_{-\infty}^{\infty}d^2{\bf r} e^{-\frac{{\bf r}^2}{2w^2}}\frac{({\bf r}-{\bf r}_i)\cdot({\bf r}-{\bf r}_j)}{2\xi^2+({\bf r}-{\bf r}_i)^2} \frac{1}{2\xi^2+({\bf r}-{\bf r}_j)^2}\nonumber\\
&\equiv&2N^2n_in_j I_{8},
\label{integral-17}
\end{eqnarray}
where $N$ is the normalization factor of the trial wave function given in equation (\ref{norm}).

\begin{figure} 
\centering
\includegraphics[width=0.65\linewidth]{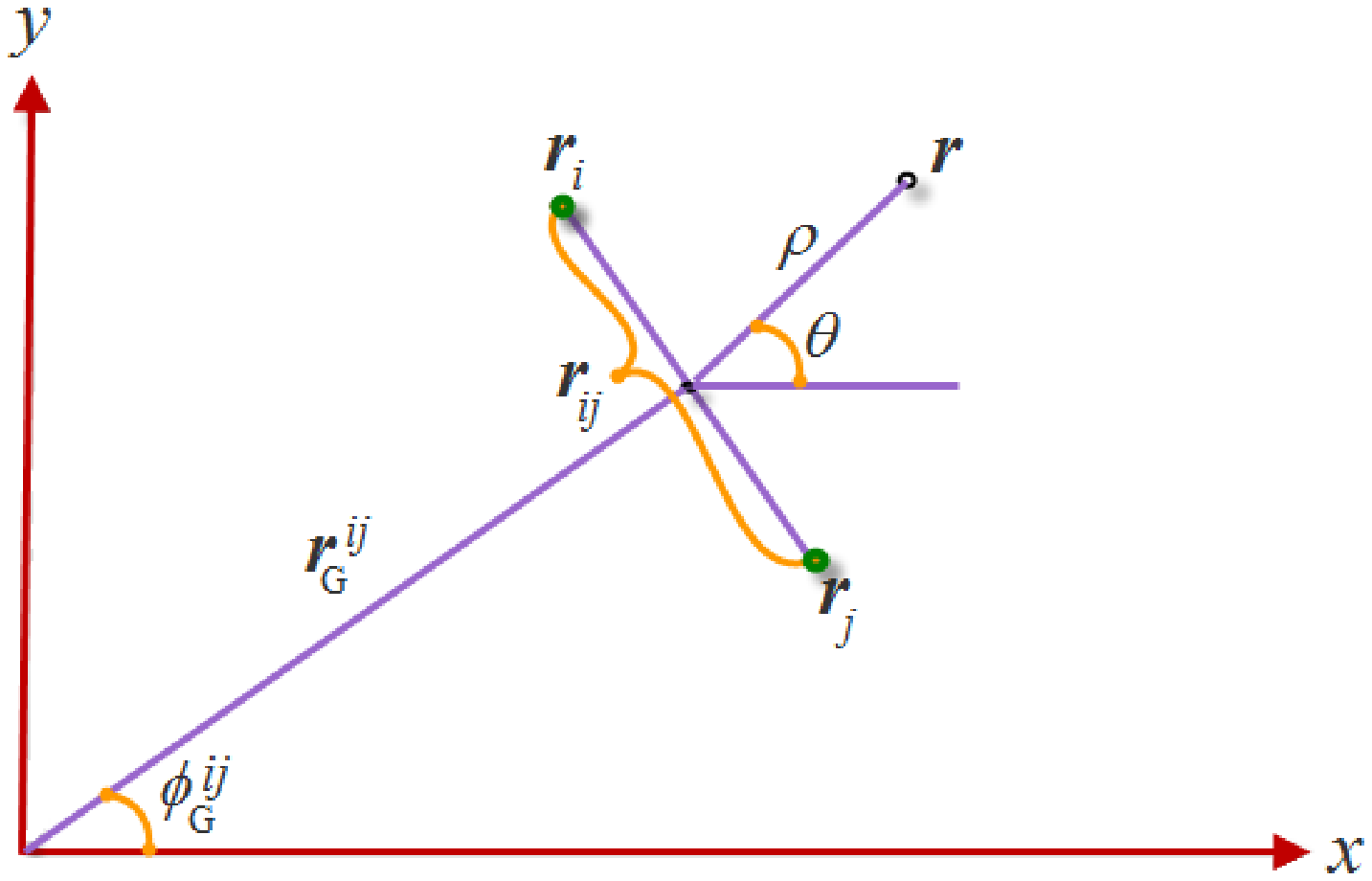}
\caption{(Color online) New integration variables: center-of-mass and relative
coordinates.} 
\label{fig:angle2}
\end{figure}

Let's define the center-of-mass and relative coordinates by
\begin{equation}\label{com cood}
x_G^{ij}=\frac{x_j+x_i}{2}; \quad y_G^{ij}=\frac{y_j+y_i}{2}
\end{equation}
and
\begin{equation}
x_i-x_j=r_{ij}\cos{\phi_{ij}}; \quad y_i-y_j=r_{ij}\sin{\phi_{ij}},
\end{equation}
respectively, and transform the integration variables to new ones as (see figure~\ref{fig:angle2})
\begin{eqnarray}
x-x_G^{ij}&=&\rho\cos{\theta}, \quad x_G^{ij}=r_G^{ij}\cos{\phi_G^{ij}}\nonumber\\
y-y_G^{ij}&=&\rho\sin{\theta}, \quad y_G^{ij}=r_G^{ij}\sin{\phi_G^{ij}}\nonumber\\
dxdy&=&\rho d\rho d\theta.
\end{eqnarray}
%\begin{equation}
%x^2+y^2=\rho^2+l_G^2+2\rho l_G\cos(\theta-\phi_G),
%\end{equation}
%\begin{equation}
%(x-x_i)^2+(y-y_i)^2=\rho^2+\frac{r_{ij}^2}{4}+\rho
%r_{ij}\cos(\theta-\phi_{ij}),
%\end{equation}
%\begin{equation}
%(x-x_j)^2+(y-y_j)^2=\rho^2+\frac{r_{ij}^2}{4}-\rho
%r_{ij}\cos(\theta-\phi_{ij}).
%\end{equation}
Then $I_{8}$ in equation (\ref{integral-17}) becomes
\begin{eqnarray}
%U&=&g_2\pi^2\Delta^2\left(\frac{N}{N_0}\right)^4\int\limits_0^{\infty}\rho
%d\rho\nonumber\\
%&\times&\int\limits_0^{2\pi}
%d\theta e^{\frac{-2}{\Delta}\left(l_G^2+\rho^2+2\rho l_G \cos{(\theta-\phi_G)}\right)}\nonumber\\
%&\times&\frac{\rho^2+\frac{r_{ij}^2}{4}+\rho
%r_{ij}\cos{(\theta-\phi_{ij})}}{2\xi^2+\rho^2+\frac{r_{ij}^2}{4}+\rho
%r_{ij}\cos{(\theta-\phi_{ij})}}\nonumber\\
%&\times&\frac{\rho^2+\frac{r_{ij}^2}{4}-\rho
%r_{ij}\cos{(\theta-\phi_{ij})}}{2\xi^2+\rho^2+\frac{r_{ij}^2}{4}-\rho
%r_{ij}\cos{(\theta-\phi_{ij})}}\nonumber\\
%r_{ij}\cos{(\theta-\phi_{ij})}}\nonumber\\
\fl I_{8}=\int\limits_0^{\infty}\rho
d\rho\int\limits_0^{2\pi}
d\theta' e^{\frac{-1}{2w^2}\left((r_G^{ij})^2+\rho^2+2\rho r_G^{ij} \cos{(\theta'+\phi_{ij}-\phi_G^{ij})}\right)}
%\nonumber\\ &\times&
\Bigg[\frac{\rho^2-\frac{r_{ij}^2}{4}}{(2\xi^2+\rho^2+\frac{r_{ij}^2}{4})^2-\rho^2 r_{ij}^2\cos{^2(\theta')}}\Bigg],\nonumber\\ 
\label{potential-polar}
\end{eqnarray}
where we moved to a new angle variable $\theta'\equiv\theta-\phi_{ij}$.
%We shall first carry out the $\theta$ integration and then $\rho$ integration. The $\theta$ integration of the first %term is
%\begin{equation}
%\int\limits_0^{2\pi}d\theta e^{-\frac{4\rho
%l_G^{ij}}{\Delta}\cos{(\theta-\phi_G^{ij})}}=\int\limits_0^{2\pi}d\tilde{\theta}
%e^{-\frac{4\rho l_G^{ij}}{\Delta}\cos{\tilde{\theta}}}=2\pi
%I_0\left(\frac{4\rho l_G^{ij}}{\Delta}\right).
%\end{equation}
%The integration of the second and third terms in equation~(\ref{potential-polar}) are identical.
Using the expansion
\begin{equation}
e^{-X \cos{(\theta'+\alpha)}}=\sum\limits_{n=-\infty}^{\infty}(-1)^n I_n(X) e^{i n\alpha}e^{i n\theta'}
\end{equation}
in equation (\ref{potential-polar}) and keeping the $n=0$ term,
the integration over the angle variable leads to:
\begin{eqnarray}\label{loren-compl}
\fl  \int\limits_0^{2\pi}d\theta'\frac{1}{(2\xi^2+\rho^2+\frac{r_{ij}^2}{4})^2-\rho^2
r_{ij}^2\frac{1+\cos{2\theta'}}{2}}\nonumber\\
=\frac{2\pi}{2\xi^2+\rho^2+\frac{r_{ij}^2}{4}}
%\nonumber\\&\times&
\frac{\left(\rho-\frac{r_{ij}}{2}\right)}{\sqrt{\left(2\xi^2+\left(\rho-\frac{r_{ij}}{2}\right)^2\right)}}
\frac{\left(\rho+\frac{r_{ij}}{2}\right)}{\sqrt{\left(2\xi^2+\left(\rho+\frac{r_{ij}}{2}\right)^2\right)}}, 
\end{eqnarray}
where we used the formula
$\int\limits_0^{2\pi}d\theta\frac{1}{c+b\cos{\theta}}=\frac{2\pi}{\sqrt{(c-b)(c+b)}}$.

We shall proceed to $\rho$ integration. Here  we should note: In the case $\xi\ll1$,
the second factor in the final result in equation~(\ref{loren-compl}) shows a step-function-like behavior around $\rho=\pm \frac{r_{ij}}{2}$ and  is well approximated for $\rho>0$ by $1-2f(\rho)$, where $f(\rho)$ is Fermi-Dirac type function defined by $\frac{1}{\exp(\frac{\rho-\frac{r_{ij}}{2}}{\xi})+1}$.
Therefore,
the $\rho$ integration leads to:
\begin{eqnarray}
I_{8}&=&2\pi e^{-\frac{(r_G^{ij})^2}{2w^2}}
%\nonumber\\&\times&
\Bigg[\int\limits_0^{\infty}\frac{\rho
e^{-\frac{\rho^2}{2w^2}}}{2\xi^2+\rho^2+\frac{r_{ij}^2}{4}}
I_0\left(\frac{\rho r_G^{ij}}{w^2}\right)\left(1-2f(\rho)\right)
d\rho\Bigg]\nonumber\\
&\approx&2\pi
e^{-\frac{(r_G^{ij})^2}{2w^2}}\Bigg[\Big[\left(\Gamma
(0,\frac{r_{ij}^2}{4w^2})-\frac{1}{2}\Gamma
(0,\frac{r_{ij}^2}{8w^2})\right)
%\nonumber\\&\times& 
\exp (\frac{r_{ij}^2}{8w^2})
\Big]+\frac{(r_G^{ij})^2}{4w}\frac{\partial}{\partial
w}\Big[\cdots\Big]\Bigg].\nonumber \\
\end{eqnarray}
where we used the approximation $I_0\left(\frac{\rho
r_G^{ij}}{w^2}\right)\approx1+\frac{(r_G^{ij})^2}{4w^4}\rho^2$.
$\Gamma(0,z)\equiv\int\limits_z^\infty t^{-1} e^{-t}dt$ is the
incomplete gamma function of the second kind and its expansion
with respect to $z(>0)$ is given by
\begin{equation}\label{gamma-in}
\Gamma(0,z)=-\gamma-\ln{z}+z-\frac{z^2}{4}+O(z^3).
\end{equation}

Finally, in the asymptotic region $\xi\ll 1$ and
$\frac{r_{ij}}{w}\ll 1$, with use of the expansion in
equation~(\ref{gamma-in}) we reach
\begin{eqnarray}\label{U-final}
I_{8}&=&-\pi \left(1-\frac{(r_G^{ij})^2}{2w^2} +\frac{r_{ij}^2}{8w^2}\right)
%\nonumber\\&\times&
\left(\gamma+\ln\left(\frac{r_{ij}^2}{2w^2}\right)-
\frac{4(r_G^{ij})^2+3r_{ij}^2}{8w^2} \right).
\end{eqnarray}
$I_8$ together with $N^2=\frac{1}{2\pi w^2}(1+O((\frac{\xi}{w})^2))$ in equation~(\ref{norm})
determines $V_{tp}$ in equation~(\ref{integral-17}). As seen in equations (\ref{v-inter}) and (\ref{magni}),  $U_{tp}\equiv  
-w^2V_{tp}$ gives
a charge-dependent part of the inter-vortex interaction in equation~(\ref{magni}).

\section{Breathing mode without vortices}\label{brm-no vortex}

In this Appendix, we consider the case without vortices and show that our scheme for condensates  gives the well-known breathing mode.
Although our main interest lies in the two-dimensional system, we consider the breathing modes of width dynamics in one, two, and three-dimensional systems.
In the $d$-dimensional system, the trial function without vortices becomes
\begin{equation}
\Phi_d(t,{\bf r}) = N_d e^{-\frac{r^2}{4 w^2} + i r^2 \beta},
\end{equation}
with the normalization factor
\begin{eqnarray}
%\begin{split}
N_1 &=& \frac{1}{(2 \pi w^2)^{1/4}}, \nonumber \\
N_2 &=& \frac{1}{\sqrt{2 \pi w^2}}, \nonumber \\
N_3 &=& \frac{1}{(8 \pi^3 w^6)^{1/4}}.
%\end{split}
\end{eqnarray}
The Lagrangian density $\mathcal{L}_d$ obtained from equation~(\ref{eq-lagrangian-density}) becomes
\begin{eqnarray}
\fl \mathcal{L}_1 = - \frac{1}{8 \pi w^5} \big[ 2 g w^3 e^{- r^2 / w^2}  + \sqrt{2 \pi} r^2 \{ 1 + 4 w^4 (1 + 4 \beta^2 + \dot{\beta})\} e^{-r^2 / (2 w^2)} \big],\nonumber \\
\fl \mathcal{L}_2 = - \frac{1}{8 \pi^2 w^6} \big[ g w^2 e^{- r^2 / w^2}  + \pi r^2 \{ 1 + 4 w^4 (1 + 4 \beta^2 + \dot{\beta})\} e^{-r^2 / (2 w^2)} \big], \nonumber\\
\fl \mathcal{L}_3 = - \frac{1}{16 \pi^3 w^7} \big[ g w e^{- r^2 / w^2}  + \sqrt{2 \pi^3} r^2 \{ 1 + 4 w^4 (1 + 4 \beta^2 + \dot{\beta})\} e^{-r^2 / (2 w^2)} \big].
\end{eqnarray}
Integrating the Lagrangian density over the space coordinate $\bf{r}$, we obtain the Lagrangian $L_d$ as
\begin{eqnarray}
%\begin{split}
L_1 &=& - \frac{g w + \sqrt{\pi} \{ 1 + 4 w^4 (1 + 4 \beta^2 + \dot{\beta})\} }{4 \sqrt{\pi} w^2}, \nonumber \\
L_2 &=& - \frac{g + 4 \pi \{ 1 + 4 w^4 (1 + 4 \beta^2 + \dot{\beta})\} }{8 \pi w^2}, \nonumber\\
L_3 &=& - \frac{g + 12 \pi^{3/2} w \{ 1 + 4 w^4 (1 + 4 \beta^2 + \dot{\beta}) \}}{16 \pi^{3/2} w^3}.
%\end{split}
\end{eqnarray}
From the Euler-Lagrange equation in equation~(\ref{eq-width}), we obtain the continuous equation
\begin{equation}
\beta = \frac{\dot{w}}{4 w},
\end{equation}
for $\beta$, and the dynamical equation
\begin{eqnarray}
%\begin{split}
 \frac{g w}{\sqrt{\pi}} + 2 & = 8 w^4 (1 + 4 \beta^2 + \dot{\beta})  \quad &  (d = 1), \nonumber\\
 \frac{g}{4 \pi} + 1 & = 4 w^4 (1 + 4 \beta^2 + \dot{\beta}) \quad & (d = 2), \nonumber\\
 \frac{g}{8 \pi^{3/2}} + w & = 4 w^5 (1 + 4 \beta^2 + \dot{\beta}) \quad &  (d = 3),
%\end{split}
\end{eqnarray}
for $w$.
Here, we consider the collective dynamics around the stationary state $w = w_0$ and $\beta = 0$.
$w_0$ satisfy
\begin{eqnarray}
%\begin{split}
 \frac{g w_0}{\sqrt{\pi}} + 2&  = 8 w_0^4 \quad & (d = 1), \nonumber\\
\frac{g}{4 \pi} + 1 & = 4 w_0^4 \quad &(d = 2), \nonumber\\
\frac{g}{8 \pi^{3/2}} + w_0 & = 4 w_0^5 \quad &(d = 3),
%\end{split}
\end{eqnarray}
In the limit of $g \to 0$ and $g \gg 1$, $w_0$ becomes $w_0 \to 1 / \sqrt{2}$ and $w_0 \to g^{1 / (2 + d)} / (2 \pi^{d / (4 + 2 d)})$ respectively.
We divide $w$ to the stationary value $w_0$ and the fluctuation $\delta w$ as $w = w_0 + \delta w$, obtaining the continuous equation
\begin{equation}
\dot{\delta w} = 4 \beta w_0 + O((\delta w,\beta)^2)
\end{equation}
and the dynamical equation
\begin{eqnarray}
%\begin{split}
\frac{g \delta w}{\sqrt{\pi}} & = 32 w_0^3 \delta w + 8 w_0^4 \dot{\beta} + O((\delta w, \beta)^2) \quad &  (d = 1), \nonumber\\
0 & = 4 w_0^3 \delta w + w_0^4 \dot{\beta} + O((\delta w, \beta)^2) \quad & (d = 2), \nonumber\\
\delta w & = 20 w_0^4 \delta w + 4 w_0^5 \dot{\beta} + O((\delta w, \beta)^2) \quad & (d = 3),
%\end{split}
\end{eqnarray}
within the linear order of $w_0$ and $\delta w$.
These equations gives the breathing dynamics of BEC with frequencies
\begin{eqnarray}
%\begin{split}
 \omega_1^2 & = 16 - \frac{g}{2 \sqrt{\pi} w_0^3} \quad & (d = 1),\nonumber \\
 \omega_2^2 & = 16 \quad & (d = 2), \nonumber\\
\omega_3^2 & = 20 - \frac{1}{w_0^4} \quad & (d = 3),
\label{TF-bm}
%\end{split}
\end{eqnarray}
where $\omega_d$ is the frequency of the breathing mode in the $d$-dimensional system.
In the limiting values of $g$, equation~(\ref{TF-bm}) accords with
the  known results in the Thomas-Fermi approximation \cite{Stri,Jin,Mew,fett1,fett2}:
for $g \to 0$, $\omega_d \to 4$ (irrespective of $d$), and for $g \to \infty$, 
\begin{eqnarray}
%\begin{split}
& \omega_1 \to 2 \sqrt{3} \quad & (d = 1), \nonumber \\
& \omega_2 \to 4 \quad & (d = 2), \nonumber \\
& \omega_3 \to 2 \sqrt{5} \quad & (d = 3),
%\end{split}
\end{eqnarray}
respectively.

%!!!!!!!!!!!!!!!!!!!!!!!!!!!!!!!!!!!!!!!!!!!!!!!!!!!!!!!!!!!!!!!!!!!!!!!!!!!!!!!!!!!!!!!!!!!!!!!!!!!!!!!!!!!!!!!!!!!!!!!!!!!!!!!!!!!!!!!!!!!!!!!!!!!!!!!!!!!!!!!!!!!!!!!!!!!!!!!!!!!!!!!!!!!!!!!!!!!!!!!!!!!!!!!!!!!!!!!!
\end{document}